# FIRST PERFORMANCE RESULTS OF THE PIP2IT MEBT 200 OHM KICKER PROTOTYPE*

G. Saewert[†], M.H. Awida, B.E. Chase, A. Chen, J. Einstein-Curtis, D. Frolov, K. Martin, H. Pfeffer, D. Wolff, Fermilab, Batavia 60510, USA
S. Khole, BARC, India
D. Sharma, RRCAT, India

*Abstract*

The PIP-II project is a program to upgrade the Fermilab accelerator complex. The PIP-II linac includes a 2.1 MeV Medium Energy Beam Transport (MEBT) section that incorporates a unique chopping system to perform arbitrary, bunch-by-bunch removal of 162.5 MHz structured beam. The MEBT chopping system will consist of two identical kickers working together and a beam absorber. One design of two having been proposed has been a 200 Ohm characteristic impedance traveling wave dual-helix kicker driven with custom designed high-speed switches. This paper reports on the first performance results of one prototype kicker built, installed and tested with beam at the PIP-II Injector Test (PIP2IT) facility. The helix deflector design details are discussed. The electrical performance of the high-speed switch driver operating at 500 V bias is presented. Tests performed were chopping beam at 81.25 MHz for microseconds as well as with a truly arbitrary pattern for 550 μs bursts having a 45 MHz average switching rate and repeating at 20 Hz.

## INTRODUCTION

The PIP-II MEBT Chopper scheme is described in [1,2,3] along with early kicker development efforts. The development of a switch-based kicker driver and 200 Ω characteristic impedance kicker version was proposed because of the lower power demanded of the driver compared to a typical 50 Ω system, the advent of GaN FETs meant potentially a 500 to 600 V switch could be built to switch fast enough to kick individual bunches, and a switch would be DC coupled to the kicker eliminating settling issues when abruptly changing to different arbitrary kicking patterns or duty factors.

## SYSTEM DESIGN

Figure 1 shows the kicker overall design. Design efforts focused on: helix design, driver design, obtaining a vacuum feed through imposing minimum reflections, and obtaining a wide-band load also with low reflections.

### Purchased Components

The feed throughs were custom-made by MPF Products, Inc., Gray Court, SC, 29645. The design, although targeted to be 200 Ω cylindrical, is physically short enough that it appears as a capacitor < 1 pF. Both center conductor and cylindrical return are electrically isolated from the flange.

The load was prototyped by Elab, Inc., elabinc@mtnhome.com, with resistive elements supplied by Smiths Interconnect, emc-rflabs.com.

### 200 Ohm Helical Deflector

Table 1 lists major helix characteristics. Early pre-prototypes revealed a 200 Ω helix could have bandwidth adequately beyond 500 MHz. But dispersion-type effects and end reflections resulting from sub-two nanosecond rise times are pronounced and had to be reduced. RF modeling the helix [4] became useful to help understand and optimize the design only after correlating results with tests of a structure having the same geometry and analyzing the software model the same way that time domain measurements were made on the bench.

The modelling results were used to make final dimensional choices for this version. End reflections result, because the helix is a microstrip line whose conductors are close enough together to couple and increase the characteristic impedance by 40%. The coupling is missing on the last turn, so the impedance is low, thus the reflections. Modelling showed reducing the tube diameter under the last turn helped reduce reflected voltage peak by ~30%.

Table 1: Dual-helix physical dimensions

| Parameter | Value |
| --- | --- |
| Helix length | 501.4 mm |
| Helix pitch | 10.46 mm |
| Number of turns | 47.5 |
| Aperture | 13 mm |
| Wire size, #13 Cu | 2.67 x 1.07 mm |
| AlN ceramic height | 6.10 mm |
| AlN thermal cond. | 50 W/mK |
| Ground tube dia. | 28.35 mm |
| Cu electrode dim. | 7.62 W x 20 L x 0.050 Th mm |
| Velocity factor | 0.0667 |

Dimensions were chosen for phase velocity to be ~0.5% faster and slightly higher impedance than our 185 Ω target to be able to make fine adjustments after construction. Both these parameters are reduced in value by inserting a dielectric – for example Kapton – the full helix length under the windings in each quadrant. However, measurements show increasing helix group delay start at several hundred megahertz. And compensating this by adding

---

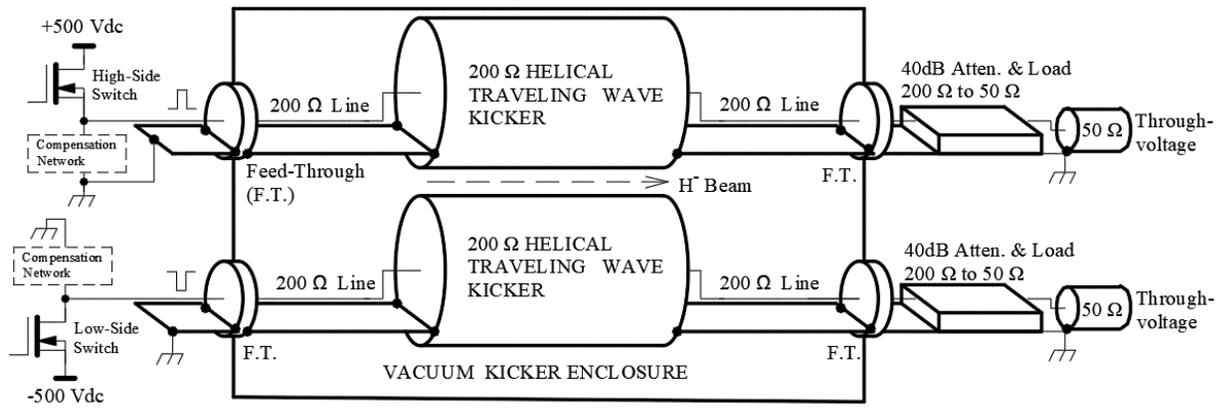

Figure 1: Overall 200 Ω kicker system. Ground connections are soldered to the vacuum feed throughs. System bandwidth at > 600 MHz enables high resolution pulse response monitoring at "through-voltage" outputs.

small turn-to-turn coupling capacitance decreased the otherwise excessive dispersion. But this also lowered the propagation velocity. This compensation was applied to both helices rather than inserting dielectric.

The RF conductor losses in the 200 Ω helix are estimated to be less than 8 W, but to also handle 40 W direct beam loss, the ground tube is water cooled. Machinable AlN ceramic spacers support the helical wire at each quadrant around the ground tube, and to enhance thermal conduction epoxy is applied between all copper surfaces and the ceramic. Thermal measurements revealed wire temperature rises < 10 ºC in the vacuum when the wire was made to dissipate 40 W.

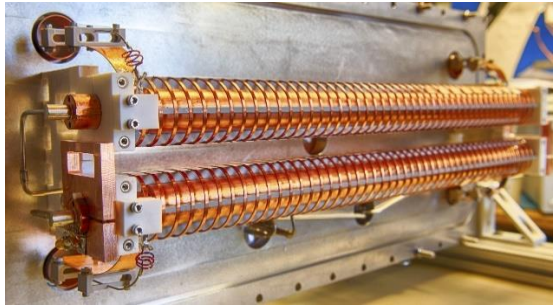

Figure 2: Dual-helix beam deflector. At the near-end are helices outputs, 200 Ω strip-lines and feed throughs.

*High Speed Switch Driver*

The bunch-by-bunch kicking requirements include the following (assuming the driver for one helix is a single switch):

(1) Operating voltage from 0 V to at least + or − 500 V;
(2) Transition times less than 4.0 ns , 5-95%;
(3) Pulse widths ranging from 4.0 ns FWHM, to kick or pass a single bunch, to microseconds;
(4) Drive an arbitrary pulse pattern at any duty factor removing up to 80% of the beam;
(5) Burst switching rates averaging 45 MHz for 550 µs at 20 Hz to deliver beam to the Booster;
(6) Designed in a way to be "CW compatible", meaning capable of switching continuously in the megahertz range or higher. An example would be to pass 162.5 MHz beam for ~200 ns every ~1.5 µs continuously to a Mu2e experiment.

The challenges for this kicker driver include: fast rise/fall times, generating narrow pulses and switching high repetition rates either for bursts or sustained intervals. High switching rates result in switching losses in the FETs. This is addressed by choosing a 200 Ω instead the typical 50 Ω, building the switch with multiple FETs in series to decrease individual dissipated power and using GaN instead of MOSFETs. Junction temperature rise at high switching rates result in shifting the FET's gate threshold voltage that alters on/off timing from the tuned values and unbalances voltage and power sharing among the 3 FETs in the switch.

The GaN FET driver circuit is discretely built, since there is no fast enough commercial gate driver available, yet. Each FET is driven on and off individually simultaneously, and each FET driver circuit is isolated from each other and from ground. Power is delivered to each FET driver circuit by an isolated AC power delivery system.

Development of this kicker driver scheme was reported on and discussed [5]. As mentioned, trigger transformers delivering signals to the gate driver circuits produced erratic triggering when switching above 40 MHz due to parasitic ringing on the secondaries. The present design resolves these problems by replacing the transformer circuitry with a photonics trigger system. The GaN FET driver circuit was redesigned to include a photodetector. A simple diagram of this system is shown in Fig. 3. The high and low-side switches are built with three FETs connected in series. Photos of a FET driver circuit PCB and one assembled switch are in Fig. 4.

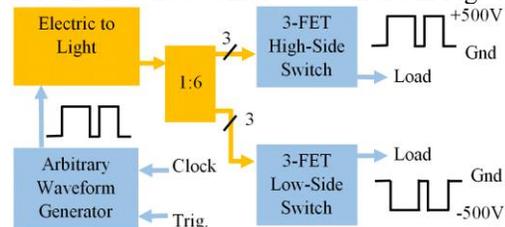

Figure 3: Kicker driver system. The Pulse pattern is delivered over fiber-optic cable to each FET driver circuit.

Depicted in Fig. 3 is a Chase Scientific arbitrary waveform generator, DAx22000, equipped with code and a LabVIEW interface written at Fermilab for describing arbitrary waveforms whose edges can be easily adjusted with ~40 ps resolution for optimum synchronization with beam bunches.

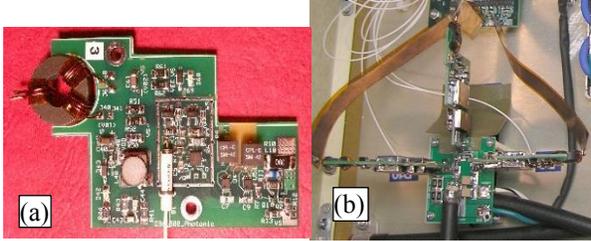

Figure 4: (a) FET gate driver circuit. (b) 3-FET switch assembly. Strips deliver AC power to gate driver circuits. Switch output drives a 185 Ω Coax.

## PERFORMANCE

One complete 200 Ω kicker system was installed into the PIP2IT facility and beam-chopping tests were performed [6]. One test was to switch at 81.25 MHz to kick out every other bunch for bursts lasting 12 µs. Beam emittance and extinction ratio were evaluated. A second test was to demonstrate chopping beam to simulate delivery into Booster with the required arbitrary waveform pattern. Chopping bursts last 550 µs with a 45 MHz average switching rate and repeated at 20 Hz. The FET Junction temperature rise above ambient switching for Booster injection have been determined to be 45 ºC during the 550 µs. This is still tolerable with forced air cooling the FETs that are mounted on G10 boards. Figures 5 – 8 demonstrate kicker voltages and respective resistive wall current monitor (RWCM) beam signals.

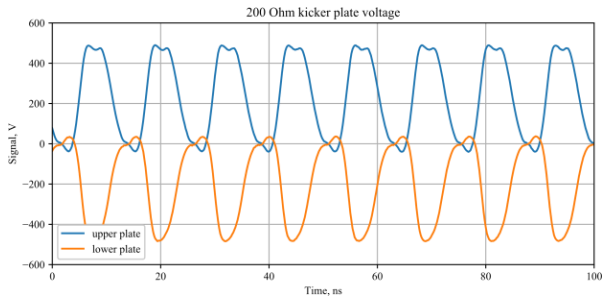

Figure 5: 81.25 MHz switching. Deflector differential voltage is on the right, the actual burst lasts 12 µs.

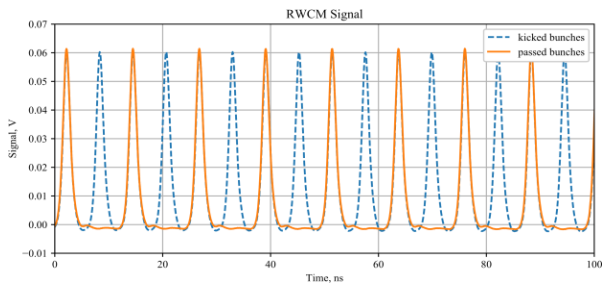

Figure 6: RWCM signal when switching 81.25 MHz switching.

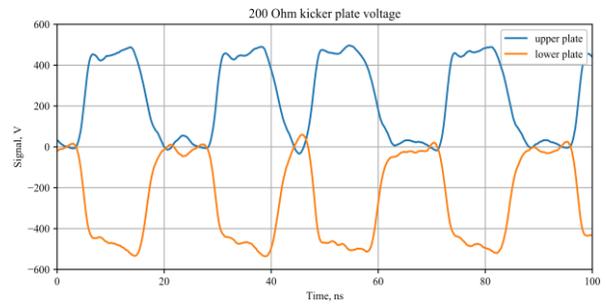

Figure 7: Booster chopping waveform with 45 MHz arbitrary pattern, actual burst lasts 550 µs.

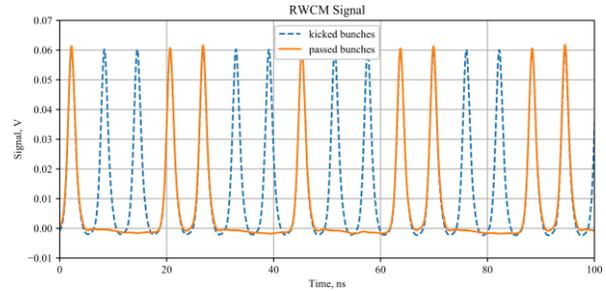

Figure 8: RWCM signal when switching Booster chopping waveform.

Also, a switch operated continuously without failure for four months switching with the Booster injection pattern demonstrating these GaN FETs were not overstressed.

## CONCLUSION

The 200 Ω kicker prototype demonstrated beam chopping acceptably for PIP-II to deliver beam into the Fermilab Booster.

## ACKNOWLEDGMENTS

The authors wish to acknowledge the dedication and efforts of Jeff Simmons for technical support and Kevin Roon for his helix winding fixture.